\documentclass[lettersize,journal]{IEEEtran}
\usepackage{amsmath,amsfonts}
\usepackage{algorithmic}
\usepackage{algorithm}
\usepackage{array}
\usepackage[caption=false,font=normalsize,labelfont=sf,textfont=sf]{subfig}
\usepackage{textcomp}
\usepackage{stfloats}
\usepackage{url}
\usepackage{verbatim}
\usepackage{graphicx}
\usepackage{cite}
\usepackage[percent]{overpic}
\usepackage{siunitx}
\usepackage{soul}
\usepackage[normalem]{ulem}

\begin{document}

\title{A Versatile Analytical Model for Fast and Accurate Determination of Feedline Coupled Resonators for Superconducting Qubit Readout}

\author{Zhen Luo,~\IEEEmembership{Graduate Student,~IEEE,}
        Léa Richard,
        Ivan Tsitsilin,
        Christian M.F. Schneider,
        Marco Dietz,\\
        Stefan Filipp and
        Amelie Hagelauer,~\IEEEmembership{Senior Member,~IEEE,} \\
\thanks{This work was supported in part by the Federal Ministry of Education and Research (BMBF) of Germany as part of Munich Quantum Valley Quantum Computer Demonstrators for Superconducting Qubits (MUNIQC SC) under Project 13N16188, and in part by Munich Quantum Valley (MQV) supported by the Bavarian state government with funds from the Hightech Agenda Bayern Plus, and in part by grants from the EU MSCA Cofund International, Interdisciplinary, and Intersectoral Doctoral Program in Quantum Science and Technologies (QUSTEC) (GrantNr. 847471).}
\thanks{Zhen Luo is with the Technical University of Munich, TUM School of Computation, Information and Technology, Chair of Micro- and Nanosystems Technology, 85748 Garching bei München, Germany (e-mail: zhen.luo@tum.de).}
\thanks{Léa Richard, Ivan Tsitsilin, Christian Schneider and Stefan Filipp are with the Technical University of Munich, TUM School of Natural Sciences, Department of Physics,
85748 Garching bei München, Germany, and also with Walther-Meißner-Institut, Bayerische Akademie der Wissenschaften, 85748 Garching bei München, Germany (e-mail: lea.richard@wmi.badw.de).}
\thanks{Marco Dietz is with the Fraunhofer EMFT Institute for Electronic
Microsystems and Solid State Technologies, 80686 München, Germany.}
\thanks{Amelie Hagelauer is with the Technical University of Munich, TUM School of Computation, Information and Technology, Chair of Micro- and Nanosystems Technology, 85748 Garching bei München, Germany, and also with the Fraunhofer EMFT Institute for Electronic Microsystems and Solid State Technologies, 80686 München, Germany (e-mail: amelie.hagelauer@tum.de).}
}


\maketitle

\begin{abstract}

Superconducting quantum chips commonly utilize quarter-wavelength ($\lambda/4$) transmission line resonators as readout circuits. An analytical model for the accurate determination of resonance frequencies and coupling Q-factors of feedline-coupled superconducting resonators is introduced.
The model leverages four-port microwave network analysis, integrating boundary conditions and conformal mapping techniques to compute even- and odd-mode impedances in edge-coupled coplanar waveguide (CPW) structures. Its versatility allows application to both planar and three-dimensional heterogeneous architectures, making it a powerful tool for resonator design. To validate the model, a test chip with $\lambda/4$ resonators of varying geometries is fabricated and measured in a cryogenic environment. Comparisons with finite element method (FEM) simulations and experimental measurements confirm the model's accuracy, with resonance frequencies and coupling Q-factors aligning closely across configurations. This proposed model facilitates the design of superconducting resonators in readout circuits for more effective, scalable, and adaptable quantum computing architectures. 

\end{abstract}

\begin{IEEEkeywords}
Coplanar waveguide, conformal mapping, coupled-line resonator, readout circuit, 3D heterogeneous integration, superconducting quantum chip
\end{IEEEkeywords}

%
\IEEEpeerreviewmaketitle


\section{Introduction}

\IEEEPARstart{T}{he} increasing number of quantum-bits (qubits) on superconducting quantum chips sets higher and higher demands on the architecture and routing of control and readout signals \cite{Conner_2021}. Early systems primarily use a planar architecture with single metal layers \cite{Kelly_2015, Song_2017}. However, as the number of qubits rises, the limitations of the planar architecture with a single metal layer become prominent. The most important among these constraints are the routing challenge for control and readout of qubit states, and the challenge of managing unwanted interactions between qubits and nearby components on an increasingly crowded chip \cite{Rosenberg_2017}. Air-bridge technology \cite{Chen_2014} has been proposed as an interim solution to avoid line crossings for routing \cite{Mukai_2020}. However, a more promising solution is to employ flip-chip technology, which introduces an additional metal layer on a top substrate die, leveraging the third dimension \cite{Rosenberg_2017, Norris_2024}. This additional metal layer is stacked on top of the bottom die with indium bumps to provide interconnects \cite{foxen2017}. This architecture not only alleviates routing congestion, but also provides effective shielding to minimize crosstalk among on-chip elements \cite{Kosen_2024}, \cite{Barrett_2023}, \cite{Dai_2021}.

The readout resonators in quantum circuits are typically designed as meandering quarter-wavelength ($\lambda/4$) transmission line resonators coupled to a feed line, a configuration well suited for dispersive readout techniques \cite{Blais2004, Krantz_2019}. Achieving the desired performance requires precise control over both the resonance frequency ($f_r$) and the coupling quality factor ($Q_c$). For the conventional planar architecture, the model presented in \cite{Besedin_2018} provides effective tools for design and analysis. However, this model is specifically optimized for planar systems and does not account for the additional complexities introduced by the top metal layer present in the flip-chip architecture. 

To address these limitations, another model proposed in \cite{Li_2023} provides accurate estimations of both $f_r$ and $Q_c$ for configurations with a top ground plane. This model leverages 2D cross-sectional simulations using a finite element method (FEM) electromagnetic (EM) solver to evaluate the coupling between the resonator and feedline. While this approach accelerates the design process for $\lambda/4$ resonators and significantly reduces computational resources compared to full-wave 3D FEM simulations for parameter determination, a numerical model with closed-form expressions would be even more advantageous for rapid resonator synthesis. Additionally, for circuits with an additive resonator-to-qubit coupling pad, which is typically placed at the open end of the $\lambda/4$ resonator to provide sufficient capacitive coupling between the readout resonator and qubit, this model requires additional curve-fitting steps. These steps rely on adequate 3D FEM simulations to collect sufficient statistics for accurate fitting, ensuring that the model achieves the desired level of precision. However, this approach is only valid for round-shaped coupling pads, which limits the model’s applicability to other design choices.


\begin{figure*}[htbp]
  \centering
  \begin{overpic}[width=\textwidth]{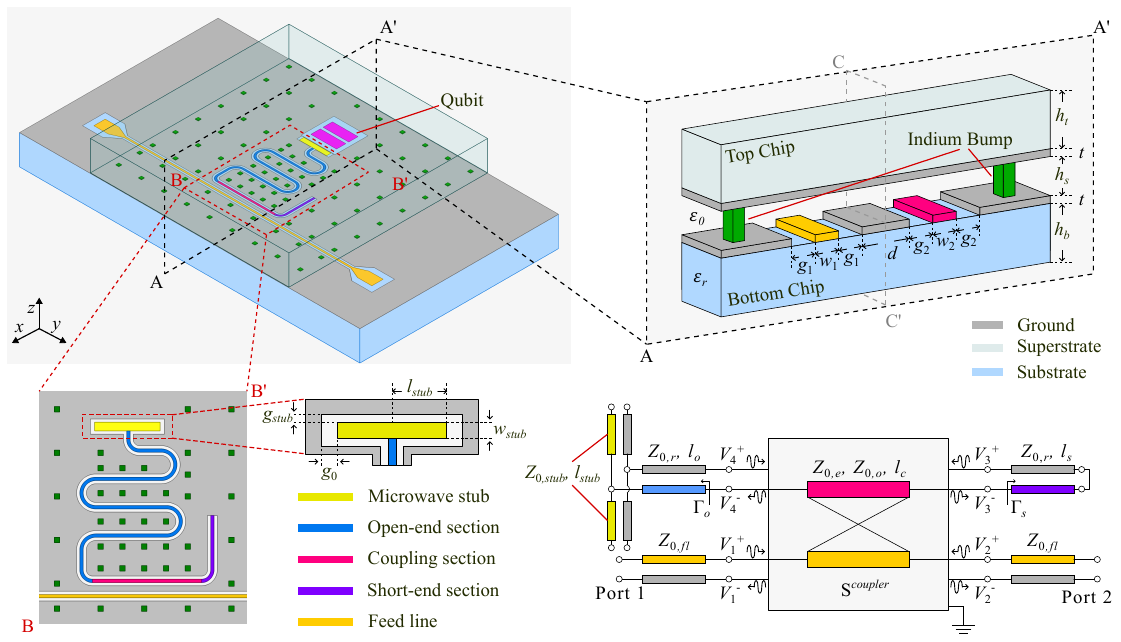}
  \put(1, 55){\normalsize (a)}
  \put(52, 55){\normalsize (b)}
  \put(1, 22){\normalsize (c)}
  \put(52, 22){\normalsize (d)}
  \end{overpic}
  \caption{(a) False-color isometric view of the 3D model, illustrating the feedline-coupled readout resonator system in a flip-chip configuration. The qubit and readout resonator are located on the bottom chip, while the top chip is fully metallized except for some apertures above the qubit. The top chip is intentionally diced to a smaller size than the bottom chip, allowing additional space for wire-bonding pads along the feedline. (b) Cross-sectional view of the coupling section with a false scale to illustrate the stack-up. (c) Top view of the physical layout of the resonator with the top chip unveiled. The resonator is divided into four sections: short-end section (purple), coupling section (magenta), open-end section (blue), and microwave stub (yellow). (d) Distributed element-based circuit model of the system with corresponding colors as in (c).}
  \label{fig:main}
\end{figure*}
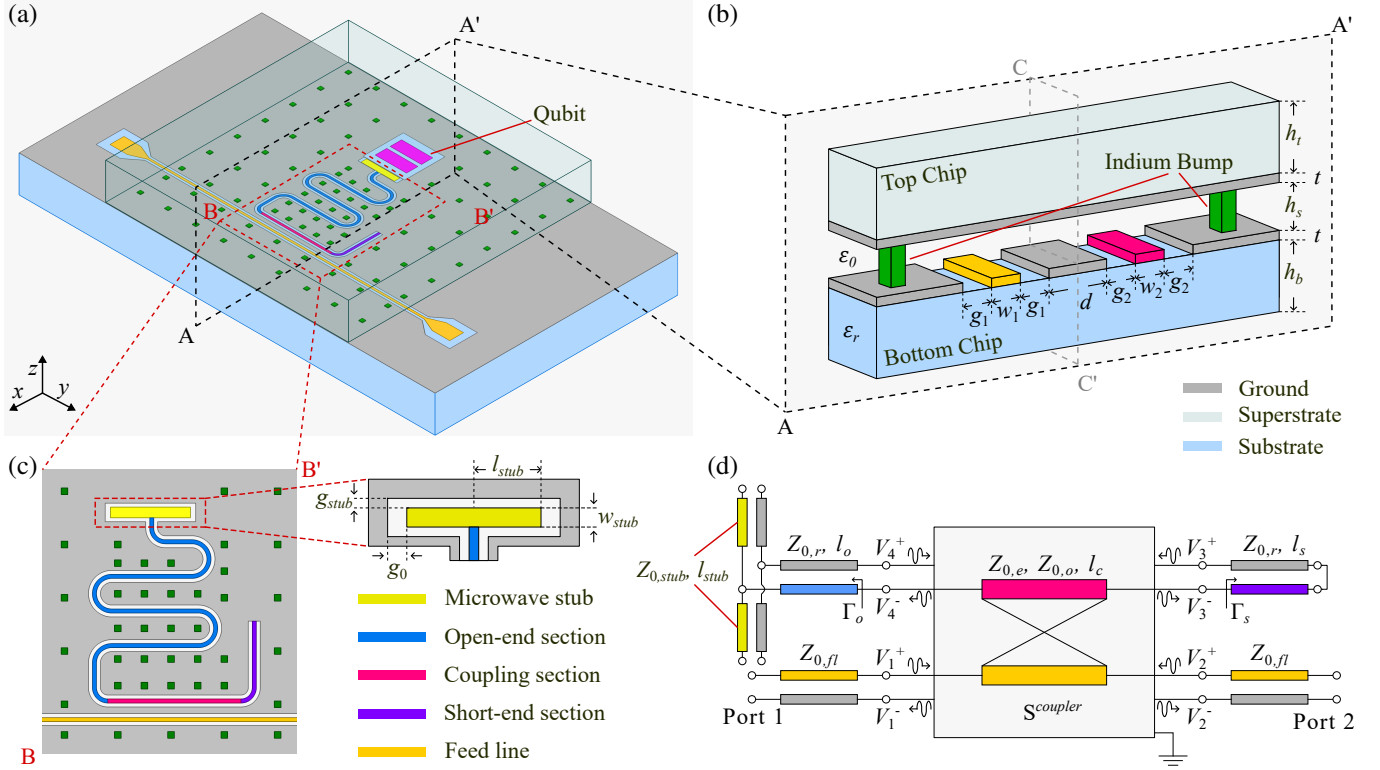

In this paper, we develop closed-form expressions to characterize the feedline-coupled resonator system. The analysis of the coupling section is based on the even- and odd-mode approach combined with conformal mapping techniques. The proposed model represents the system as a two-port microwave network, enabling a concise expression for the $S_{21}$ transmission spectrum. By inputting only key geometric parameters — such as the distance between the resonator and feedline, the resonator's total length, and the location of the coupling section, etc.—the model enables rapid calculation of the transmission coefficient $S_{21}$. This approach facilitates efficient extraction of $f_r$ and $Q_c$, simplifying the design and analysis of quantum readout circuits. Furthermore, the proposed model includes the coupling pad as an additive microwave stub in the circuit, accurately capturing its effects on the system. Another notable advantage of the proposed model is its versatility across various stack-ups. It can be applied to both conventional planar structures and 3D heterogeneous architectures that incorporate a lower metal layer on the bottom die \cite{Rosenberg_2017}. With minor modifications, the model also supports configurations with a broadside-coupling scheme between the resonator and the feedline, making it adaptable to diverse design requirements.

We validate the model by comparing its predictions to 3D full-wave Ansys HFSS simulations for resonators with varying geometric parameters. Additionally, we design and fabricate a test chip utilizing flip-chip technology, with a spacing of \qty{10}{\micro m} between the upper and lower dies. Key geometric parameters are varied across different designs, and measurements are conducted within a cryogenic setup to establish a benchmark. The primary parameters obtained from the model, $f_r$ and $Q_c$, agree well with simulations and experimental data, with discrepancies mainly due to fabrication tolerances and spurious modes existing in the coupling section. These results demonstrate that the proposed model provides a reliable tool for designing readout circuits for large-scale qubit chips, significantly reducing simulation time and improving design efficiency. 

\section{Analytical Model}
\label{sect:2}
In this section, we present a generic analytical model for determining the eigenmode resonant frequency ($f_r$) and the coupling Q-factor ($Q_c$) of a feedline-coupled $\lambda/4$ transmission line resonator. Fig.~\ref{fig:main}(a) illustrates an isometric view of a transmon qubit \cite{Koch_2007} with a pocket ground and its readout circuit. The feedline and resonator, implemented using a CPW configuration, are separated by a finite ground plane with a lateral width $d$. Normally, the feedline and resonator have identical geometric parameters, such that $w_1=w_2=w$ and $g_1=g_2=g$. The bottom chip faces a fully metallized layer on the flipped chip, with a chip separation $h_s$ between the two chips. Indium bumps, placed ubiquitously with a pitch size \qty{50}{\micro m}, provide a galvanic interconnection between the top and bottom chips. The detailed stack-up configuration is shown in Fig.~\ref{fig:main}(b). 

The readout resonator structure consists of four parts: a short-end section, a coupling section, an open-end section, and an ending pad, as shown in Fig.~\ref{fig:main}(c). The distributed element based model for each of these sections is shown in Fig.~\ref{fig:main}(d).

\subsection {Edge-coupled CPW Line Coupler}

From the definition of the impedance matrix $\mathbf{Z}$, the four-port network of the coupling section, as represented by the ivory-colored box in Fig.~\ref{fig:main}(d), can be expressed in terms of the voltages and currents at each port:
\begin{equation}\label{coupler_z_matrix}
\begin{bmatrix}
    V_{1} \\
    \vdots\\
    V_{4}
\end{bmatrix}
=
\begin{bmatrix}
    Z_{11} & ... & Z_{14} \\
    \vdots & \ddots &\vdots\\
    Z_{41} & ... & Z_{44}    
\end{bmatrix}
\begin{bmatrix}
    I_{1} \\
    \vdots\\
    I_{4} \\
\end{bmatrix}.
\end{equation}

\begin{figure}[t]
    \centering
    \begin{overpic}{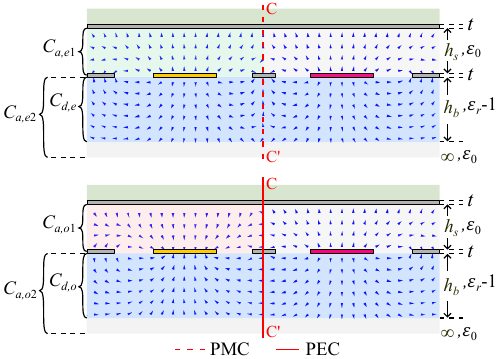} 
        \put(0, 68){\normalsize (a)}  
        \put(0, 33){\normalsize (b)}
    \end{overpic}
    \caption{E-field distributions for the configuration of top-grounded edge-coupled CPW line with a finite ground plane simulated using Ansys 2D Extractor and the configuration of partial capacitances for the even mode (a) and the odd mode (b).}
    \label{fig:field_distribution}
\end{figure}

For two adjacent parallel transmission lines, two fundamental propagation modes are supported: the even mode and the odd mode, depending on the phase of the signal. The corresponding electrostatic field distributions for these modes are shown in Fig~\ref{fig:field_distribution}, obtained using the Q2D Extractor in Ansys Electronics Desktop. The entries of the impedance matrix in (\ref{coupler_z_matrix}) can be determined through even- and odd-mode analysis \cite{Jensen2008}:
\begin{equation}\label{z_entries}
\begin{aligned}
    Z_{11}=Z_{22}=Z_{33}=Z_{44} =  \frac{-j}{2}(Z_{0,e} + Z_{0,o}) \cot\theta \\
    Z_{12}=Z_{21}=Z_{34}=Z_{43} =  \frac{-j}{2}(Z_{0,e} - Z_{0,o}) \cot\theta \\
    Z_{13}=Z_{31}=Z_{24}=Z_{42} =  \frac{-j}{2}(Z_{0,e} - Z_{0,o}) \csc\theta \\
    Z_{14}=Z_{41}=Z_{23}=Z_{32} =  \frac{-j}{2}(Z_{0,e} + Z_{0,o}) \csc\theta \\
\end{aligned}
\end{equation}
where $\theta=\beta l_c$ is the electrical length of the coupling section and $\beta$ is the wavenumber, while $Z_{0,e}$ and $Z_{0,o}$ denote the characteristic impedances of the even and odd modes, respectively.

Since both the even and odd modes are quasi-TEM modes, their characteristic impedances are primarily determined by the cross-sectional geometry of the coupled structure. As a result, quasi-static conformal mapping techniques can be applied to derive closed-form analytical expressions for the effective dielectric constant ($\varepsilon_{\mathrm{eff},e}$ and $\varepsilon_{\mathrm{eff},o}$) and the characteristic impedance ($Z_{0,e}$ and $Z_{0,o}$). Due to symmetry, a virtual Perfect Magnetic Conductor (PMC) or Perfect Electrical Conductor (PEC) boundary can be assigned along the symmetric plane ($\mathrm{CC'}$ in Fig.~\ref{fig:field_distribution}) for the even mode and odd mode, respectively, without disturbing the field distribution. This symmetry reduces the computational domain, allowing the analysis to be restricted to only half of the structure. Note that, for the following analysis, the superconductor thickness $t$ is assumed to be zero. While this assumption may introduce some discrepancy, it remains reasonable given the typical parameter values in a CPW structure, where $t \ll w$.

The cross-section of the top-grounded coupled CPW line with a finite ground plane can be divided into three partial regions \cite{Veyres1980ExtensionOT}, with the partial capacitances contributed from each region denoted separately in Fig.~\ref{fig:field_distribution}. The total capacitance of the structure is the sum of the partial capacitances:
\begin{equation}\label{C_total}
C_{tot,m} = C_{a,m1} + C_{a,m2} + C_{d,m},
\end{equation}
where $m$ represents the mode, either even ($e$) or odd ($o$). Here, the terms $C_{a,m1}$, $C_{a,m2}$ and $C_{d,m}$ denote the partial capacitances per unit length of different regions: $C_{a,m1}$ corresponds to the air region within the chip separation cavity, $C_{a,m2}$ is the partial capacitance of the lower region in the absence of the dielectric substrate and $C_{d,m}$ is the partial capacitance of the lower dielectric region. Fig.~\ref{fig:field_distribution} shows the configuration of all partial capacitances in even and odd excitations. 

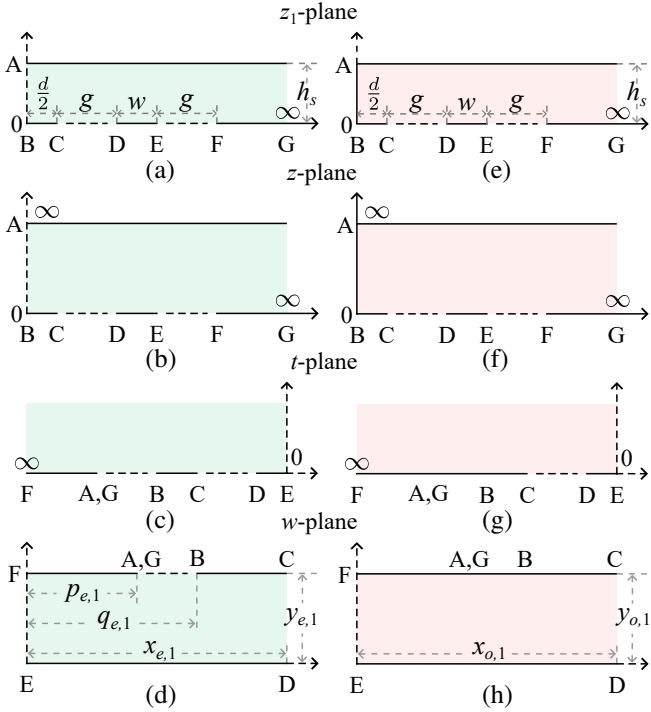
\begin{figure}[t]
    \centering
    \begin{overpic}{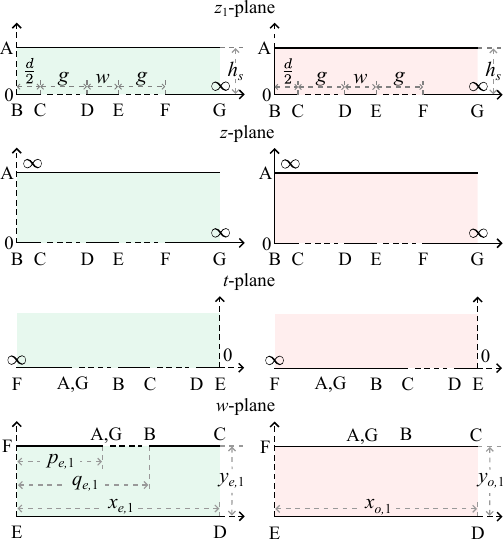} 
        \put(20, 75){\normalsize (a)}  
        \put(20, 48){\normalsize (b)}
        \put(20, 25){\normalsize (c)}
        \put(20, 0){\normalsize (d)}
        \put(68, 75){\normalsize (e)}
        \put(68, 48){\normalsize (f)}
        \put(68, 25){\normalsize (g)}
        \put(68, 0){\normalsize (h)}
        
    \end{overpic}    
    \caption{Conformal mapping transformation steps to calculate the partial capacitance $C_{a,m1}$ for even mode (a)-(d) and odd mode (e)-(h). The solid lines represent PEC boundaries and the dash lines represent PMC boundaries.}
    \label{fig:conformal_mapping}
\end{figure}

Through the sequences of transformation steps in Fig.~\ref{fig:conformal_mapping}, half of the air region in chip separation cavity,  depicted as the green and red regions in Fig.~\ref{fig:field_distribution}(a) and (b), is mapped into a parallel-plate structure to evaluate $C_{a, e1}$ and $C_{a, o1}$, respectively. To achieve this, the PEC boundary provided by the top ground with a finite distance, $h_s$, from the structures on the bottom chip must first be transformed to a layer with an infinite distance from the bottom chip. This transformation requires an additional preparatory step to map the $z_1$-plane to the $z$-plane using:
\begin{equation}\label{z1_to_z}
    z = \sinh(\pi z_1/2h).
\end{equation}
The geometries in $z_1$-plane with a boundary condition at finite distance, $h_s$, defined by the coordinates:
\begin{equation}\label{point_in_z}
    \begin{aligned}
        z_c &= d/2, \\
        z_d &= d/2 + g, \\
        z_e &= d/2 + g + w,\\
        z_f &= d/2 + 2g + w, \\
    \end{aligned}
\end{equation}
are subsequently transformed into corresponding points in the $z$-plane by applying (4):
\begin{equation}\label{zs_points}
    z_{s,n} = \sinh(\pi z_n/2h_s), \quad n=c, d, e, f.
\end{equation}
For both the even and odd modes, the parallel-plane structures obtained after transformation have the same width and height as:
\begin{equation}\label{ko1}
\begin{gathered}
    x_{e1} = x_{o1} = K(k_{o1}), \\
    y_{e1} = y_{o1} = K'(k_{o1}), \\
    k_{o1} = \sqrt{\frac{(z_{s,f}^2 - z_{s,c}^2)(z_{s,e}^2 - z_{s,d}^2)}{(z_{s,e}^2 - z_{s,c}^2)(z_{s,f}^2 - z_{s,d}^2)}},
\end{gathered}
\end{equation}
where $K(k)$ and $K'(k)$ are the complete elliptic integrals of the first kind. The relationship between these integrals is given by:
\begin{equation}\label{K(k)}
        K'(k) = K(k'),
        k' = \sqrt{1-k^2}.
\end{equation}


It is important to note that the parallel-plate structure for the even mode, after transformation, features an aperture notch on one plate, while the transformed parallel-plate structure for the odd mode consists of two complete plates. The position of this notch in $w$-plane is defined by the value of $p_{e1}$ and $q_{e1}$ in Fig.~\ref{fig:conformal_mapping}(d), and is determined using:
\begin{equation}\label{pe1_qe1}
    \begin{gathered}
        p_{e1} = F\left(\arcsin\sqrt{\frac{z_{s,c}^2(z_{s,f}^2-z_{s,d}^2)}{z_{s,d}^2(z_{s,f}^2-z_{s,c}^2)}}, k_{o1}\right), \\
        q_{e1} = F\left(\arcsin\sqrt{\frac{\cosh^2(\pi z_c/2h_s)(z_{s,f}^2-z_{s,d}^2)}{\cosh^2(\pi z_d/2h_s)(z_{s,f}^2-z_{s,c}^2)}}, k_{o1}\right),
    \end{gathered}
\end{equation}
where $F(\varphi, k)$ is the incomplete elliptic integral of the first kind. The partial capacitance $C_{a,e1}$ and $C_{a,o1}$ are then calculated using \cite{Cheng1997}:
\begin{equation}\label{C_ae1_Cao1}
    \begin{gathered}
        C_{a,e1} = \varepsilon_0C_p\left(\frac{x_{e1}}{y_{e1}}, \frac{p_{e1}}{x_{e1}}, \frac{q_{e1}}{x_{e1}}\right), \\
        C_{a, o1} = \varepsilon_0 \frac{K(k_{o1})}{K'(k_{o1})}, \\
    \end{gathered}
\end{equation}
where $C_p$ is a function of three geometric inputs, $\alpha,\beta,\gamma$, which can be computed using \cite{Cheng1995}: 
\begin{equation}\label{Cp}
    \begin{gathered}
    C_p(\alpha,\beta,\gamma) = K(k_1)/K(k_1') + K(k_3)/K(k_3'), \\
    \delta = \frac{(\beta+\gamma)}{2}, \\
    \frac{K(k_4)}{K(k_4')} = \alpha(1 - \delta), \\
    \frac{K(k_2)}{K(k_2')} = \alpha\delta, \\
    \frac{F[\arcsin(k_3/k_4), k_4]}{K(k_4)} = \frac{1-\gamma}{1-\delta}, \\
    \frac{F[\arcsin(k_1/k_2), k_2]}{K(k_2)} = \frac{\beta}{\delta}.
    \end{gathered}
\end{equation}
Some useful formulas for solving (\ref{pe1_qe1}) and (\ref{Cp}) are provided in \cite{Cheng1995}. 

For the partial capacitance in the lower region, $C_{a,m2}$ depicted in Fig.~\ref{fig:field_distribution}, the air region boundary at the bottom is at an infinite distance from the structures. Consequently, the transformation from $z_1$-plane to $z$-plane can be omitted, and the conformal mapping results are directly reshaped as:
\begin{equation}\label{ko2}
\begin{gathered}
    x_{e2} = x_{o2} = K(k_{o2}), \\
    y_{e2} = y_{o2} = K'(k_{o2}), \\
    k_{o2} = \sqrt{\frac{(z_{f}^2-z_{c}^2)(z_{e}^2-z_{d}^2)}{(z_{e}^2-z_{c}^2)(z_{f}^2-z_{d}^2)}}, \\
\end{gathered}    
\end{equation}
where 
\begin{align*}
    k_{o2} = \sqrt{\frac{(z_{f}^2-z_{c}^2)(z_{e}^2-z_{d}^2)}{(z_{e}^2-z_{c}^2)(z_{f}^2-z_{d}^2)}.}
\end{align*}
The position of the notch for the even mode after transformation is changed to
\begin{equation}\label{ko2}
\begin{gathered}
    p_{e2} = F\left(\arcsin\sqrt{\frac{z_{c}^2(z_{f}^2-z_{d}^2)}{z_{d}^2(z_{f}^2-z_{c}^2)}}, k_{o2}\right), \\
    q_{e2} = F\left(\arcsin\sqrt{\frac{z_{f}^2-z_{d}^2}{z_{f}^2-z_{c}^2}}, k_{o2}\right).
\end{gathered}    
\end{equation}
Finally, the partial capacitance, $C_{a,m2}$, can also be computed by
\begin{equation}\label{C_ae2_Coe2}
    \begin{gathered}
        C_{a,e2} = \varepsilon_0C_p\left(\frac{x_{e2}}{y_{e2}}, \frac{p_{e2}}{x_{e2}}, \frac{q_{e2}}{x_{e2}}\right), \\
        C_{a, o2} = \varepsilon_0 \frac{K(k_{o2})}{K'(k_{o2})}. \\        
    \end{gathered}
\end{equation}

When evaluating the partial capacitance $C_{d,e}$ and $C_{d,o}$, the finite thickness of the substrate ($h_b$) must be taken into account. To address this, a transformation step is applied to convert the finite boundary to an infinite one. The transformed positions for the dielectric layers with infinite thickness can be obtained by replacing $h_s$ in (\ref{z1_to_z}) by $h_b$ as:
\begin{equation}\label{zb_points}
    z_{b,n} = \sinh(\pi z_n/2h_b), n=c, d, e, f.
\end{equation}
Similarly the partial capacitance $C_{d,m}$ are computed by the following equations:
\begin{equation}\label{C_de_Cdo}
    \begin{gathered}
        C_{d, e} = \varepsilon_0C_p\left(\frac{x_{d}}{y_{d}}, \frac{p_{d}}{x_{d}}, \frac{q_{d}}{x_{d}}\right), \\
        C_{d, o} = \varepsilon_0 \frac{K(k_{o3})}{K'(k_{o3})}, \\      
    \end{gathered}
\end{equation}
where 
\begin{align*}
        & k_{o3} = \sqrt{\frac{(z_{b,f}^2-z_{b,c}^2)(z_{b,e}^2-z_{b,d}^2)}{(z_{b,e}^2-z_{b,c}^2)(z_{b,f}^2-z_{b,d}^2)}}, \\
        & p_{d} = F\left(\arcsin\sqrt{\frac{z_{b,c}^2(z_{b,f}^2-z_{b,d}^2)}{z_{b,d}^2(z_{b,f}^2-z_{b,c}^2)}}, k_{o3}\right), \\
        & q_{d} = F\left(\arcsin\sqrt{\frac{(z_{b,f}^2-z_{b,d}^2)}{(z_{b,f}^2-z_{b,c}^2)}}, k_{o3}\right). \\
\end{align*}

At this point, all the terms in (\ref{C_total}) have been determined using closed-form expressions. Consequently, the effective dielectric constants for both the even and odd modes, $\varepsilon_{\mathrm{eff}, m}$, and the characteristic impedances are calculated using:
\begin{gather}
\varepsilon_{\mathrm{eff}, m} = \frac{C_{tot,m}}{C_{a,m}} = \frac{C_{tot,m}}{C_{a,m1}+C_{a,m2}}, \label{eps_eff}\\
Z_{0,m} = \frac{1}{c_0 C_{a,m} \sqrt{\varepsilon_{\mathrm{eff}, m}}}, \label{Z_0m}
\end{gather}
where $c_0$ is the speed of light in free space. By solving $Z_{0,e}$ and $Z_{0,o}$ and substituting the results into (\ref{z_entries}), the entries for the impedance matrix of the coupled CPW line with a finite ground plane in the flip-chip configuration, $\mathbf{Z}^{\text{coupler}}$, can be obtained. Furthermore, the corresponding scattering matrix of the coupler, $\mathbf{S}^{\text{coupler}}$, can be converted using
\begin{equation}\label{s_matrix}
\mathbf{S}^{\text{coupler}} = \left(\mathbf{Y}_0 \cdot \mathbf{Z}^{\text{coupler}} - \mathbf{Z}_0\right) \cdot \left(\mathbf{Y}_0 \cdot \mathbf{Z}^{\text{coupler}} + \mathbf{Z}_0 \right)^{-1},
\end{equation}
where $\mathbf{Z_0}$ is a diagonal matrix having the square root of the characteristic impedance at each port $\mathbf{Z_0}=\mathrm{diag}\left[\sqrt{Z_{01}},...,\sqrt{Z_{04}}\right]$ and $\mathbf{Y_0}=\mathbf{Z_0}^{-1}$.

\subsection {Circuit Model for Feedline Coupled Resonator}
The circuit in Fig.~\ref{fig:main}(d) shows an example that the feedline is connected to port 1 and port 2 of the coupler, while the short-end and open-end sections of the $\lambda/4$ resonator are connected to port 3 and port 4, respectively. In this configuration, the reflection coefficients $\Gamma_{s}$ and $\Gamma_{o}$, seen looking toward the short- and open-end of the resonator, are determined by the lengths of the respective sections, $l_s$ and $l_o$:
\begin{align}
    \Gamma_{s} &= \frac{jZ_{0,r}\tan(\beta l_{s}) - Z_{0}}{jZ_{r}\tan(\beta l_{s}) + Z_{0}},\label{gamma_s} \\
    \Gamma_{o} &= \frac{-jZ_{0,r}\cot(\beta l_{o}) - Z_{0}}{-jZ_{r}\cot(\beta l_{o}) + Z_{0}}, \label{gamma_o}
\end{align}
where $Z_{0,r}$ is the transmission line characteristic impedance of the resonator, which is determined by the geometry of the CPW structure, and $Z_0$ is the port characteristic impedance. The effective dielectric constant, $\varepsilon_{\mathrm{eff}, r}$, and characteristic impedance, $Z_r$, for the top-grounded CPW line can be calculated using conformal mapping techniques with following equations \cite{WANG1990}:
\begin{gather}
\varepsilon_{\mathrm{eff}, r} = 1+(\varepsilon_r-1)\frac{K(k_{r,3})}{K'(k_{r,3})}\frac{1}{\frac{K(k_{r,1})}{K'(k_{r,1})}+\frac{K(k_{r,2})}{K'(k_{r,2})}}, \label{eps_cpw}\\
Z_r = \frac{1}{c\varepsilon_0\sqrt{\varepsilon_{\mathrm{eff}, r}}\left(\frac{K(k_{r,1})}{K'(k_{r,1})}+\frac{K(k_{r,2})}{K'(k_{r,2})}\right)}, \label{Z0_cpw}
\end{gather}
where
\begin{align*}
    k_{r,1} &= w/(w+2g), \\
    k_{r,2} &= \tanh(\pi w/4h_s)/\tanh(\pi (w+2g)/4h_s), \\
    k_{r,3} &= \sinh(\pi w/4h_b)/\sinh(\pi (w+2g)/4h_b).
\end{align*}
Since the feedline is also a top-grounded CPW line with identical geometry, its characteristic impedance, $Z_{0,fl}$, can be determined using the same approach as in (\ref{Z0_cpw}). Typically, the characteristic impedances in the system are designed to be identical, such that $Z_{0,r}=Z_{0,fl}=Z_0=50\Omega$. Under this condition, the reflection coefficients in (\ref{gamma_s}) and (\ref{gamma_o}) are simplified to $\Gamma_s=-1e^{-j2\beta l_s}$ and $\Gamma_o=1e^{-j2\beta l_o}$. Subsequently, based on the definition of scattering matrix and the fact that $V_{3}^{+} = \Gamma_{s}V_{3}^{-}$ and $V_{4}^{+} = \Gamma_{o}V_{4}^{-}$, the outgoing waves on port 3 and port 4 are given by:
\begin{gather}
     V_{3}^{-} = S_{31}V_{1}^{+} + S_{32}V_{2}^{+} + S_{33}\Gamma_{s}V_{3}^{-} + S_{34}\Gamma_{o}V_{4}^{-}, \label{V3-}\\
     V_{4}^{-} = S_{41}V_{1}^{+} + S_{42}V_{2}^{+} + S_{43}\Gamma_{s}V_{3}^{-} + S_{44}\Gamma_{o}V_{4}^{-}, \label{V4-}
\end{gather}
where $S_{ij}$ are the entries in the S-matrix of the coupler $\mathbf{S}^{\text{coupler}}$ solved in (\ref{s_matrix}). 

In transmission measurement, the outgoing wave at port 2, $V_2^-$ is measured by applying an incident wave at port 1, $V_1^+$. Under the determined short- and open-boundary conditions at port 3 and port 4, imposed by the resonator, the four-port network, $\mathbf{S}^{\text{coupler}}$, is reduced to a two-port network, denoted by a newly formed scattering parameter $\mathbf{S}'$. The transmission response of $S_{21}'$ along the feedline is used to characterize the resonator and is defined in relation to the incident and reflected voltage waves as:
\begin{equation}\label{S21}
    S_{21}' = \frac{V_{2}^{-}}{V_{1}^{+}} \bigg|_{V_{2}^{+} = 0} = S_{21}+S_{23}\Gamma_{s}V_{3}^{-} + S_{24}\Gamma_{o}V_{4}^{-}.
\end{equation}
Eliminating $V_4^-$ in (\ref{V3-}) and under the condition that $V_2^+=0$, $V_3^-$ can be expressed in terms of $V_1^+$:
\begin{equation}\label{V3_full}
 \begin{aligned}
    V_{3}^{-} = \frac{1}{A} (S_{31} + \frac{S_{34}S_{41}\Gamma_{o}}{1-S_{44}\Gamma_{o}})V_{1}^{+}, \\
    A = 1-S_{33}\Gamma_{s} - \frac{S_{34}S_{43}\Gamma_{o}\Gamma_{s}}{1-S_{44}\Gamma_{o}}. \\
 \end{aligned}
\end{equation}
Similarly, eliminating $V_3^-$ in (\ref{V4-}), $V_4^-$ can be also expressed in terms of $V_1^+$: 
\begin{equation}\label{V4_full}
\begin{aligned}
V_{4}^{-} = \frac{1}{B} (S_{41} + \frac{S_{43}S_{31}\Gamma_{s}}{1-S_{33}\Gamma_{s}})V_{1}^{+}, \\
B = 1-S_{44}\Gamma_{o} - \frac{S_{34}S_{43}\Gamma_{o}\Gamma_{s}}{1-S_{33}\Gamma_{s}}. \\    
\end{aligned}
\end{equation}
Subsequently, by inserting (\ref{V3_full}) and (\ref{V4_full}) into (\ref{S21}), a general model for the transmission coefficient from port 1 to port 2, $S_{21}'$, of the reduced two-port network is derived as:
\begin{equation}\label{s21_full}
\begin{split}
    S_{21}{'} =&\: S_{21} + \frac{S_{23}\Gamma_{s}}{A}(S_{31} + \frac{S_{34}S_{41}\Gamma_{o}}{1-S_{44}\Gamma_{o}})  \\
    & + \frac{S_{24}\Gamma_{o}}{B}(S_{41} + \frac{S_{43}S_{31}\Gamma_{s}}{1-S_{33}\Gamma_{s}}).
\end{split}
\end{equation}
Here, $S_{ij}$ are the entries of $\mathbf{S}^{\text{coupler}}$, which are determined by the geometric parameters of the coupling section, as discussed in the previous analysis. The reflections, $\Gamma_s$ and $\Gamma_o$, are determined by the geometric parameters of the short- and open-end section of the $\lambda/4$ resonator. Consequently, a closed-form expression for the transmission coefficient of an edge-coupled resonator, in relation to its geometry, is obtained. 

Since the frequency response $S_{21}'(f)$ is available in (\ref{s21_full}), the resonance frequency $f_r$ can, in principle, be determined by solving $|S_{21}'(f_r)| = 0$, and the loaded quality factor $Q_l$ can be calculated as $Q_l = 1/\text{FBW}$, where FBW is the 3dB fractional bandwidth. However, the resulting expressions are analytically cumbersome and impractical for use. Therefore, in this work, we employ a circuit fitting approach to extract the parameters, such as $f_r$ and $Q_l$, as discussed in detail in Section \ref{subsec:determin_fr_Qc}.

Note that the reverse transmission coefficient of the reduced two-port network from port 2 to port 1, $S_{12}'$, can be modeled with the same approach and is mathematically identical  to $S_{21}'$ in (\ref{s21_full}), confirming that the network is reciprocal. Similarly, the reflection coefficient at port 1, $S_{11}'$, when all other ports are terminated, can also be resolved using the same methodology and is useful for single-port measurement.  

\subsection {Modeling the Bifurcated Coupling Pad}
In typical superconducting qubit systems, the open-end of the $\lambda/4$ readout resonator is placed close to the qubit to provide capacitive coupling between resonator and qubit. To achieve sufficient coupling strength, a bifurcated ending pad is commonly added at the open-end, as shown in Fig.\ref{fig:main}(c). This ending pad can be modeled as two microwave open-circuited stubs in parallel, incorporated into the circuit depicted in Fig.\ref{fig:main}(d). 

The resonance frequency of readout resonators in superconducting qubit systems is typically chosen to be within the $6$–\qty{8}{GHz} range, while the length of the ending pad is normally on the order of several hundred micrometers. Therefore, the stub remains electrically short with respect to the wavelength, satisfying $l_{stub}<\lambda/10$. In this case, the admittance of this open-circuited stub can be approximated as:
\begin{equation}\label{Y_stub}
    Y_{\mathrm{stub}} = j\omega \left(\frac{l_{\mathrm{stub}}\sqrt{\varepsilon_{\mathrm{eff, stub}}}}{cZ_{0,\mathrm{stub}}} \right).
\end{equation}
Here, the effective dielectric constant, $\varepsilon_{\mathrm{eff, stub}}$, and the characteristic impedance, $Z_{0,\mathrm{stub}}$, of the microwave stub section can be determined by (\ref{eps_cpw}) and (\ref{Z0_cpw}). 

Subsequently, the reflection coefficient, $\Gamma_{o}^{stub}$, seen looking from the output of port 3 towards the ending pad is given by:
\begin{equation}\label{gamma_stub}
    \Gamma_o^{\mathrm{stub}} = \frac{Z_{\mathrm{stub}} - 2Z_{0,r}}{Z_{\mathrm{stub}} + 2Z_{0,r}}e^{-j2\beta l_o}.
\end{equation}
The proposed analytical model is then modified for the resonator with additive microwave stubs at the open-end position by replacing the $\Gamma_o$ in (\ref{s21_full}) with $\Gamma_o^{\mathrm{stub}}$ from (\ref{gamma_stub}).

\begin{figure}[h!]
    \centering
    \begin{overpic}{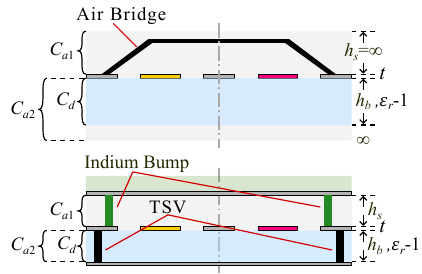} 
        \put(-10, 60){\normalsize (a)}  
        \put(-10, 25){\normalsize (b)}
    \end{overpic}
    \caption{Partial capacitances and cross-sectional view of the coupling section for planar architecture (a) and flip-chip architecture with back conductor on bottom chip (b). The plane of symmetry is designated by dash-dot line.}
    \label{fig:Different Stackup}
\end{figure}

\subsection {Model Versatility}
The proposed model shows potential extensibility to a variety of stack-up configurations. Fig.~\ref{fig:Different Stackup}(a) shows the coupling section in a planar architecture, where the circuit is implemented on a single metallization layer. In the absence of a top chip, the thickness of the upper air region is considered to be infinite. In this configuration, the partial capacitances for upper and lower air regions are identical ($C_{a,m1} = C_{a,m2}$) and can be determined using (\ref{C_ae2_Coe2}). In some envisioned stack-ups \cite{Rosenberg_2017}, an additional metallization layer can be formed on the backside of the bottom chip as shown in Fig.~\ref{fig:Different Stackup}(b). In this configuration, the partial capacitances $C_{a,m2}$ and $C_{d,m}$ are influenced by the PEC boundary condition at a finite distance $h_b$. Therefore, (\ref{zb_points}) must be applied to transform the positions of the points in the structure to $z$-plane.

Placing interconnection components such as air bridges, indium bumps, and through-silicon vias (TSVs) on the side grounds of the CPW line can eliminate voltage differences, effectively suppressing spurious modes in the coupling section. This results in an improved agreement between the proposed model and simulation results.

\begin{figure}[b]
    \centering
    \begin{overpic}{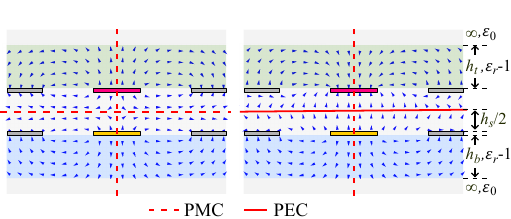} 
        \put(0, 39){\normalsize (a)}  
        \put(48, 39){\normalsize (b)}
    \end{overpic}
    \caption{Cross-sectional view of the coupling section for a broadside-coupled CPW line and the field distribution for even mode (a) and odd mode (b).}
    \label{fig:BCCPW}
\end{figure}

Additionally, the stack-up shown in Fig.~\ref{fig:main}(b) can be adapted for a broadside-coupling scheme by placing the feedline and resonator on the top and the bottom chips separately. This structure also supports even and odd modes as two fundamental propagation modes and thus can be analyzed using a similar approach to edge-coupling scheme. A virtual PMC (PEC) boundary condition can be assigned horizontally along the middle of the air cavity for even (odd) mode excitations. An additional virtual PMC boundary condition can be assigned along the plane of symmetry for both even and odd mode excitations. The field distribution profiles for the even and odd modes, simulated by Ansys Q2D Extractor, and the assignment of the virtual boundary conditions are shown in Fig.~\ref{fig:BCCPW}(a) and (b), respectively. Due to symmetry, the analysis can be restricted to one-quarter of the cross-sectional structure by leveraging quasi-static conformal mapping techniques. The effective dielectric constant and characteristic impedance for even and odd modes in the broadside-coupling configuration can be computed using the following equations \cite{Bedair1989}:
\begin{gather}
\varepsilon_{\mathrm{eff},be} = 1+(\varepsilon_r-1) \frac{K(k_{b,e3})}{K'(k_{b,e3})}\frac{1}{\frac{K(k_{b,e1})}{K'(k_{b,e1})}+\frac{K(k_{b,e2})}{K'(k_{b,e2})}}, \label{eps_e_bccpw}\\
\varepsilon_{\mathrm{eff},bo} = 1+(\varepsilon_r-1) \frac{K(k_{b,e3})}{K'(k_{b,e3})}\frac{1}{\frac{K(k_{b,e1})}{K'(k_{b,e1})}+\frac{K(k_{b,o2})}{K'(k_{b,o2})}}, \label{eps_o_bccpw}\\
Z_{0,be} = \frac{1}{c\varepsilon_0\sqrt{\varepsilon_{eff,be}} \left(\frac{K(k_{b,e1})}{K'(k_{b,e1})}+\frac{K(k_{b,e2})}{K'(k_{b,e2})}\right)}, \label{Z0_e_bccpw} \\
Z_{0,bo} = \frac{1}{c\varepsilon_0\sqrt{\varepsilon_{eff,bo}} \left(\frac{K(k_{b,e1})}{K'(k_{b,e1})}+\frac{K(k_{b,o2})}{K'(k_{b,o2})}\right)}, \label{Z0_o_bccpw}
\end{gather}
where
\begin{align*}
    k_{b,e1} &= w/(w+2g), \\
    k_{b,e2} &= \sinh(\pi w/2h_s)/\sinh(\pi (w+2g)/2h_s), \\
    k_{b,o2} &= \tanh(\pi w/2h_s)/\tanh(\pi (w+2g)/2h_s), \\
    k_{b,e3} &= \sinh(\pi w/4h_b)/\sinh(\pi (w+2g)/4h_b),\\
\end{align*}

To extend the proposed model for the broadside-coupling scheme, the entries for the four-port coupler network are obtained by substituting the calculated $Z_{0,be}$ and $Z_{0,bo}$ from (\ref{Z0_e_bccpw}) and (\ref{Z0_o_bccpw}) into (\ref{z_entries}). The analytical expressions for the transmission coefficient of the system in (\ref{s21_full}) are then updated accordingly.

\section{Model Validation}
In the experiment, \qty{525}{\micro m} bulk silicon (Si) with characterized dielectric constant $\varepsilon_r=11.45$ \cite{Krupka2006} is used as substrate. Thin-film niobium (Nb) with thickness $t=$ \qty{150}{nm} is deposited to form the metallization layers on both top and bottom chips. In simulations, the materials are defined accordingly, with PEC used to represent the superconducting Nb at cryogenic temperatures. The geometric parameters of CPW structures for both the feedline and the resonator are summarized in Table~\ref{table:cpw_geometry}.

\begin{table}[h]
\caption{Geometric Parameters of the CPW Structures and Stack-up\label{table:cpw_geometry}}
\centering
\renewcommand{\arraystretch}{1.2} 
\resizebox{0.9\columnwidth}{!}{%
\begin{tabular}{>{\centering\arraybackslash}m{0.6cm}>{\centering\arraybackslash}m{0.6cm}>{\centering\arraybackslash}m{0.6cm}>{\centering\arraybackslash}m{0.6cm}>{\centering\arraybackslash}m{0.6cm}>{\centering\arraybackslash}m{0.6cm}>{\centering\arraybackslash}m{0.6cm}>{\centering\arraybackslash}m{0.6cm}}
\hline
\hline
\text{$w_r$} & \text{$g_r$} &\text{$w_{fl}$} & \text{$g_{fl}$} & \text{$t$} & \text{$h_s$} & \text{$h_t$} & \text{$h_b$} \\
(\textmu m) & (\textmu m) & (\textmu m) & (\textmu m) & (nm) & (\textmu m) & (\textmu m) & (\textmu m) \\
\hline
\hline
10 & 9 & 10 & 9 & 150 & 10 & 525 & 525 \\ 
\hline
\end{tabular}%
}
\end{table}

\subsection {Validation of Conformal Mapping}\label{sec:conformal_mapping_verification}
\begin{figure}[t]
    \centering
    \begin{overpic}{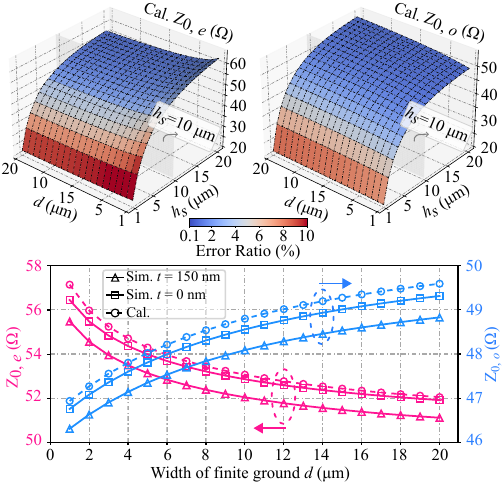} 
        \put(-2, 90){\normalsize (a)}  
        \put(49, 90){\normalsize (b)}
        \put(-2, 43){\normalsize {(c)}}
    \end{overpic}
    \caption{Comparison between calculated and simulated $Z_{0,e}$ (a) and $Z_{0,o}$ (b) of the edge coupled CPW line across varying values of $d$ and $h_s$. The error rate is depicted as a heat map. (c) Direct comparison between calculation and simulation results with swept $d$ and fixed $h_s=$ \qty{10}{\micro m}.}
    \label{fig:Sim_vs_Cal_Z0m}
\end{figure}

The coupling quality factor of the resonator is mathematically determined by the even- and odd-mode characteristic impedances ($Z_{0,e}$ and $Z_{0,o}$) in the circuit model, which are intrinsically defined by the geometric parameters in the coupling section. In particular, the lateral distance ($d$) between the resonator and feedline, as well as the vertical spacing ($h_s$) between the top and bottom chip, have significant impacts on these impedances, as calculated by the proposed conformal mapping method. The accuracy of $Z_{0,e}$ and $Z_{0,o}$ calculations has a profound influence on the overall accuracy of the model. To validate the effectiveness of the conformal mapping results, the cross-sectional structure along the $\mathrm{AA'}$-plane in Fig.\ref{fig:main}(b) is simulated using Ansys Q2D Extractor, with parameters $d$ and $h_s$ swept systematically. Fig.~\ref{fig:Sim_vs_Cal_Z0m} presents the simulation and calculation results for various values of $d$ and $h_s$. The error rate is defined as the percentage difference between simulation and calculation results, with the simulation results serving as a reference:
\begin{equation}
    \mathrm{err} = \frac{|Z_{0, \mathrm{Cal.}} - Z_{0, \mathrm{Sim.}}|}{Z_{0, \mathrm{Sim.}}}.
\end{equation}

In general, the proposed conformal mapping method provides an estimation of the characteristic impedances with an error rate below \qty{10}{\%} relative to simulation results. However, it is observed that the calculated impedances align more closely with the simulation results when $h_s>$ \qty{10}{\micro m}. This behavior arises because the conformal mapping technique assumes an infinitely thin conductor ($t=$ \qty{0}{nm}). Under this assumption, the air-dielectric interfaces in the gap between the inner conductor and the side ground in the CPW line are treated as PMC boundaries, as indicated by the dash lines in Fig.~\ref{fig:conformal_mapping}. These boundary conditions confine the electric field to be perfectly tangential between the inner conductor and ground. However, in simulations, the finite metal thickness ($t=$ \qty{150}{nm}) is explicitly modeled. The resulting discrepancy primarily arises from the fringing field caused by dispersion effects due to the finite conductor thickness. When $h_s$ is larger, the influence of finite thickness diminishes, effectively reducing the error rate. 

To further investigate the origin of the discrepancy, Fig.~\ref{fig:Sim_vs_Cal_Z0m}(c) additionally includes simulation results obtained under the condition of zero conductor thickness ($t=$ \qty{0}{nm}). As shown, when the simulation matches the model assumption of $t=$ \qty{0}{nm}, the agreement between the calculated and simulated $Z_{0,e}$ and $Z_{0,o}$ significantly improves, with an absolute error of less than \qty{0.5}{\ohm}.

Fig.~\ref{fig:Sim_vs_Cal_Z0m}(c) also presents the results from sweeping $d$ from \qty{1}{\micro m} to \qty{20}{\micro m}. The calculated values show an absolute error of less than \qty{1.5}{\ohm} from the simulation results under the condition $t=$ \qty{150}{nm} for both $Z_{0,e}$ and $Z_{0,o}$. As $d$ increases further, both impedances tend to approach the characteristic impedance of an isolated CPW line, indicating a reduction in coupling strength.

Furthermore, the results show an offset between simulation and calculation, as shown in Fig.~\ref{fig:Sim_vs_Cal_Z0m}(c). The observed offset can also be attributed to the systematic effect of finite conductor thickness. This finite thickness increases the partial capacitance in the air cavity regions $C_{a,m1}$ in (\ref{Z_0m}), thereby decreasing the characteristic impedances relative to the idealized thin-conductor case. 

\subsection {Determination of Resonance Frequency and Coupling Quality Factor}
\label{subsec:determin_fr_Qc}
The $\lambda/4$ resonator is equivalent to a parallel LCR circuit. When measuring the transmission coefficient ($S_{21}$) on the feedline to which the LCR resonator is coupled, a characteristic dip appears in the spectrum. An ideal model of the complex $S_{21}$ for such notch-type resonator, excluding environmental noise and cable attenuation, is provided in \cite{Probst_2015}:
\begin{equation}\label{S21_notch}
S_{21}^{notch}(f) = 1-\frac{Q_l/Q_c}{1+2jQ_l(f/f_r-1)}.
\end{equation}
 Here, $f_r$ represents the resonance frequency, while $Q_l$ and $Q_c$ denote the loaded and the coupling quality factors, respectively. These quantities are related by
$\frac{1}{Q_l} = \frac{1}{Q_c} + \frac{1}{Q_i}$, where $Q_i$ is the internal quality factor that accounts for intrinsic losses in the resonator. Once the transmission spectrum is obtained using (\ref{s21_full}), $f_r$ and $Q_c$ of the feedline-coupled resonator can be efficiently extracted using the curve-fitting method described in \cite{Petersan_1998}. To clarify the modeling procedure used to extract the resonance frequency and coupling quality factor, the following recursive steps summarize the pipeline proposed in this work:
\begin{enumerate}
    \item The coupled CPW line section is modeled as a four-port network. The corresponding S-parameter  ($\mathbf{S}^{\text{coupler}}$) is computed based on its physical geometry and the applied architecture, using conformal mapping technique.
    \item Short and open boundary conditions with certain distance $l_s$ and $l_o$, respectively, are applied at two of the ports to form a reduced two-port network to represent the feedline-coupled resonator circuit.
    \item The transmission coefficient $S_{21}'$ on the rest two ports is derived in closed-form expression from this network as (\ref{s21_full}).
    \item The resulting transmission response in (\ref{s21_full}) is then fitted using the standard circuit fitting procedure described in \cite{Probst_2015, Petersan_1998} to extract the resonance frequency $f_r$, coupling quality factor $Q_c$ and other parameters.
\end{enumerate}

\begin{figure}[h!]
    \centering
    \begin{overpic}{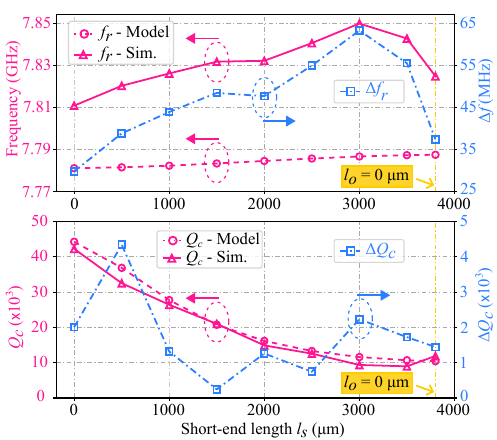} 
        \put(-2, 85){\normalsize (a)}  
        \put(-2, 43){\normalsize (b)}
    \end{overpic}
    \caption{A comparison between the calculated and the simulated $f_{r}$ (a) and $Q_{c}$ (b) of the edge coupled top grounded CPW line resonator with a fixed chip separation, $h_s=$ \qty{10}{\micro m}, and length of coupling section, $l_c=$ \qty{400}{\micro m}. An error ratio is computed in percentage respecting to the simulated results as a reference.}
    \label{fig:Sim_vs_Cal_fr_Qc}
\end{figure}

Apart from the geometric parameters $d$ and $h_s$, the coupling Q-factor is also influenced by the relative location of the coupling section on the $\lambda/4$ resonator. To study this effect and demonstrate the determination of $f_r$ and $Q_c$ using the proposed model, these parameters are calculated for a resonator with a given geometry and compared to results obtained via 3D full-wave simulation in Ansys HFSS. In this analysis, the width and gap of the CPW structure are fixed at $g_1=g_2=$ \qty{9}{\micro m} and $w_1=w_2=$ \qty{10}{\micro m}, while the geometric parameters of the coupling section are set to $d=$ \qty{5}{\micro m} and $h_s=$ \qty{10}{\micro m}. A transmission line resonator with a total length $l_t=$ \qty{4200}{\micro m} is used as an example, with microwave stubs temporarily removed for simplicity. When the length of the coupling section is fixed at $l_c=$ \qty{400}{\micro m}, the location of the coupling section can vary depending on $l_s$, where $l_t = l_s+l_c+l_o$. The results extracted from EM simulation and the proposed model for resonators with varying coupling section locations are compared in Fig.~\ref{fig:Sim_vs_Cal_fr_Qc}.

As shown in Fig.~\ref{fig:Sim_vs_Cal_fr_Qc}(a), the resonance frequencies determined by the proposed model exhibit a small discrepancy compared to simulation results. This discrepancy arises because the effective dielectric constant modeled using the conformal mapping method in (\ref{eps_cpw}) is slightly higher than the simulated value. However, the difference is within 
\qty{65}{MHz} ($\Delta f_r / f_{r,sim}<1\%$) across all variants, providing a reliable estimation. 

In contrast, the extracted $Q_c$ shows a larger discrepancy between the simulation and the model. Nevertheless, the difference remains within $5\times10^3$ ($\Delta Q_c/Q_{c,sim}<20\%$), and the trend of $Q_c$ as $l_s$ increases is consistent between the simulation and the model. This discrepancy falls within the same range as demonstrated in \cite{Li_2023} and can be attributed to three primary factors. First, spurious modes in the coupling section contribute to the reduction in $Q_c$ in simulations. While indium bumps are added along the side ground planes of the CPW line to equalize voltages, the finite ground plane between the resonator and feedline is too narrow to place indium bumps, which typically have a diameter of $10$-\qty{20}{\micro m} \cite{Kosen_2022,Lei_2020}. This lack of sufficient grounding in the narrow region introduces parasitic effects that impact the coupling quality factor. Second, discrepancies in the calculated characteristic impedances ($Z_{0,m}$) further contribute to the error. As discussed in the previous section and shown in Fig.~\ref{fig:Sim_vs_Cal_Z0m}(c), the $Z_{0,m}$ values computed using the conformal mapping method exhibit slight differences compared to simulation results, which propagate into the $Q_c$ calculations. Third, a geometric simplification in the model introduces additional errors. Specifically, in the actual resonator layout, the joint between the coupling section and the other sections of the resonator incorporates an arc-shaped connection. This arc effectively increases the length of the coupling section in the real geometry compared to the simplified right-angle transition assumed in the model. The longer effective coupling section in the actual structure results in stronger coupling and, consequently, a lower $Q_c$ in simulations. Together, these factors account for the observed differences in $Q_c$ and highlight the limitations of the model under specific conditions. However, the shift is systematic in one direction and can therefore be adjusted with an empirical parameter.

For a resonator with a predefined total length of $l_t=$ \qty{4200}{\micro m}, the resonance frequencies remain generally stable as $l_s$ increases, indicating that the location of the coupling section has a limited impact on the resonance frequency with such a $Q_c$ magnitude. In contrast, the coupling quality factor $Q_c$ is more sensitive to this parameter. In extreme cases, when $l_s=$ \qty{0}{\micro m} (or $l_o=$ \qty{0}{\micro m}), the coupling section is placed directly at the short-end (or open-end) of the resonator, resulting in predominantly inductive (or capacitive) coupling. The results suggest that capacitive coupling generally provides stronger coupling strength compared to inductive coupling. For intermediate configurations, where $l_s\not=0$ and $l_o\not=0$, a mixed-coupling scheme is supported, where both inductive and capacitive components contribute to the overall coupling.

\section{Experimental verification}

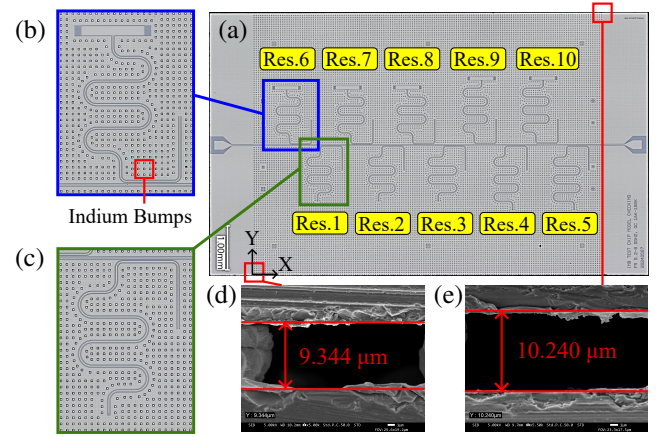
\begin{figure}[t!]
    \centering
    \begin{overpic}[width=0.92\columnwidth]{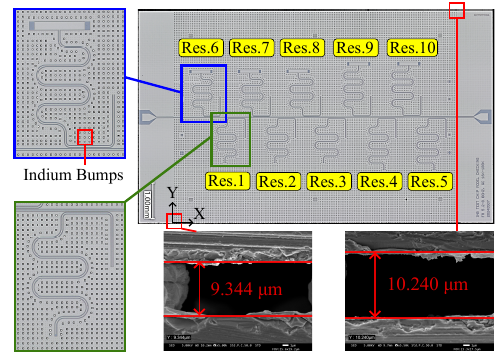} 
        \put(29, 65){\normalsize (a)}  
        \put(-4, 65){\normalsize (b)}
        \put(-4, 28){\normalsize (c)}
        \put(27, 22){\normalsize (d)}
        \put(64, 22){\normalsize (e)}
    \end{overpic}
    \caption{(a) Top view SEM picture of the test chip. (b) close-up view of the resonator with microwave stub. (c) close-up window of the typical $\lambda/4$ resonator without microwave stub. (d)-(e) side view SEM picture of the test chip to measure the value of $h_s$ at specific positions.}
    \label{fig:chip_layout}
\end{figure}

According to the previous analysis, for the flip-chip architecture with a fixed chip separation ($h_s=$ \qty{10}{\micro m}), the resonance frequency and the coupling quality factor are primarily determined by the geometric parameters of each resonator section ($l_s$, $l_o$ and $l_c$) as well as the width of the finite ground plane in the coupling section ($d$). To further validate the proposed model, a test chip comprising ten $\lambda/4$ resonators coupled to a feedline is designed, as shown in Fig.~\ref{fig:chip_layout}. Variants of the resonators, incorporating the four critical parameters mentioned above, are summarized in Table~\ref{table:geometric_parameters}. The resonators on the top half of the chip have identical geometries to those on the bottom half but include the bifurcated coupling pad on the open-end, as shown in the close-up view in Fig.~\ref{fig:chip_layout}(b). The parameters of the coupling pad are designed with parameters $l_{stub=}$ \qty{267}{\micro m}, $w_{stub}=$ \qty{80}{\micro m}, $g_{stub}=$ \qty{5.5}{\micro m} and $g_0=$ \qty{55}{\micro m}. 

\begin{table*}[!htp]
\caption{Geometric Variations of the Resonators on the Test Chip and the Corresponding Parameters $f_{r, model}/f_{r, sim}$ and $Q_{c, model}/Q_{c, sim}$ Determined by the Proposed Model and Simulations. Here, $l_c$ Is the Coupling Length, $l_s$ and $l_o$ Are the Shorted and Open-ended Section Lengths, Respectively, and $l_t = l_s + l_c + l_o$ Is the Total Resonator Length}

\centering
\renewcommand{\arraystretch}{1.5} 
\resizebox{\textwidth}{!}{%
\begin{tabular}{>{\centering\arraybackslash}p{1cm}>{\centering\arraybackslash}p{1.5cm}>{\centering\arraybackslash}p{1.5cm}>{\centering\arraybackslash}p{1.5cm}>{\centering\arraybackslash}p{1.5cm}>{\centering\arraybackslash}p{1.2cm}>{\centering\arraybackslash}p{1.5cm}>{\centering\arraybackslash}p{2cm}>{\centering\arraybackslash}p{2cm}}
\hline
\hline
\text{Index} & \text{$l_c$ (\textmu m)} & \text{$l_o$ (\textmu m)} & \text{$l_s$ (\textmu m)} & \text{$l_t$ (\textmu m)} & \text{$d$ (\textmu m)} & \text{Ending Pad} & \text{$f_r$ (GHz)} & \text{$Q_c$ ($\times 10^3$)} \\ 
\hline
\hline
Res. 1 & 400 & 3101.5 & 578.5 & 4080 & 2 & No & 8.01 / 8.05 & 17.1 / 16.4 \\ 

Res. 2 & 400 & 3316.5 & 578.5 & 4295 & 4 & No & 7.61 / 7.65 & 30.7 / 27.5 \\ 

Res. 3 & 400 & 3556.5 & 578.5 & 4535 & 6 & No & 7.21 / 7.25 & 50.7 / 45.7 \\ 

Res. 4 & 400 & 3821.5 & 578.5 & 4800 & 8 & No & 6.81 / 6.85 & 79.8 / 72.3 \\ 

Res. 5 & 400 & 4121.5 & 578.5 & 5100 & 10 & No & 6.41 / 6.45 & 122.1 / 105.6 \\ 

\hline

Res. 6 & 400 & 3101.5 & 578.5 & 4080 & 2 & Yes & 6.16 / 6.21 & 30.7 / 30.2 \\ 

Res. 7 & 400 & 3316.5 & 578.5 & 4295 & 4 & Yes & 5.92 / 5.96 & 54.1 / 43.3 \\ 

Res. 8 & 400 & 3556.5 & 578.5 & 4535 & 6 & Yes & 5.67 / 5.71 & 86.9 / 82.2 \\ 

Res. 9 & 400 & 3821.5 & 578.5 & 4800 & 8 & Yes & 5.42 / 5.45 & 132.9 / 129.3 \\ 

Res. 10 & 400 & 4121.5 & 578.5 & 5100 & 10 & Yes & 5.16 / 5.20 & 197.3 / 186.0 \\ 
\hline
\end{tabular}%
}
\label{table:geometric_parameters}
\end{table*}

\subsection {Measurement Setup}
\begin{figure}[!h]
    \centering
    \begin{overpic}{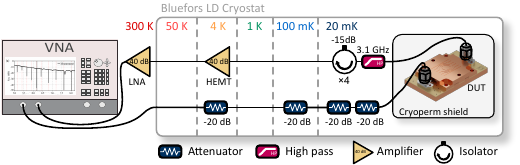} 
    \end{overpic}
    \caption{Measurement setup for cryogenic characterizations. The device under test (DUT) is wirebonded to a multi-layer PCB, magnetically shielded with a cryoperm shield, and connected to a default cryogenic wiring scheme with a total of \qty{-80}{dB} of attenuation for the input line. The output line includes a cryogenic HEMT amplifier (\textit{LNF LNC4\_8C}), 2 dual junction isolators (\textit{LNF-ISISC4\_12A}), and a \qty{3.1}{GHz} high pass filter (\textit{Minicircuits VHF-3100+}). All characterization measurements have been conducted with a vector network analyzer (\textit{Keysight N5224B}). }
    \label{fig:setup}
\end{figure}

The chip is fabricated under a flip-chip bonding process with SU-8 polymer spacers to ensure an uniform chip spacing. The fabrication process is adapted from \cite{Norris_2024}. The fabricated chip is measured in a commercial dilution fridge (\textit{Bluefors LD 400} with a base temperature of around \qty{20}{\milli\kelvin}). The measurement setup is presented in Fig.~\ref{fig:setup}.

\subsection {Analysis of Measurement Results}
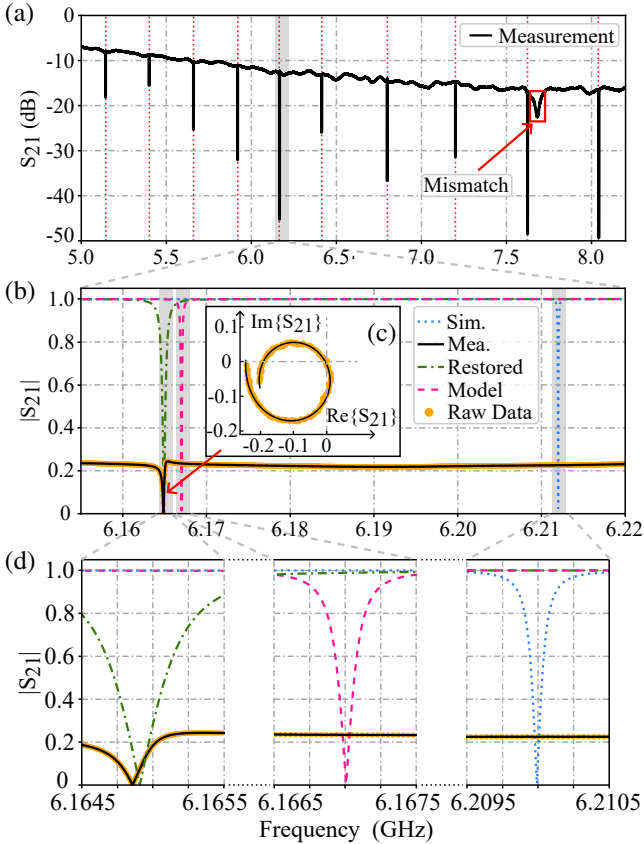
\begin{figure}[!h]
    \centering
    \begin{overpic}{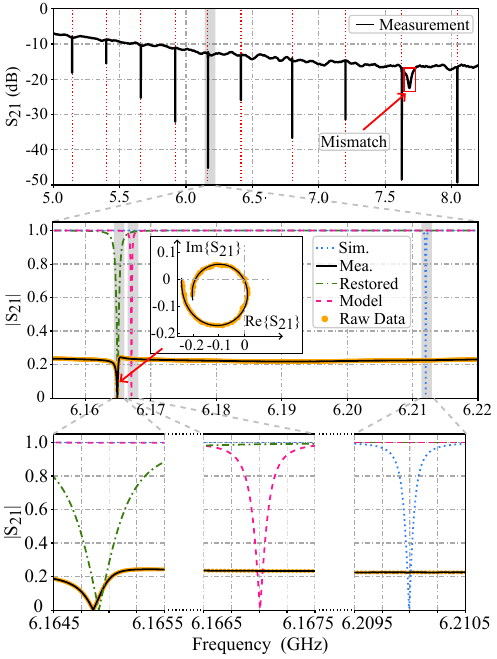} 
        \put(-1, 98){\normalsize (a)}  
        \put(-1, 65){\normalsize (b)}
        \put(42, 60){\normalsize (c)}
        \put(-1, 33){\normalsize (d)}
    \end{overpic}
    \caption{(a) Measured S21 spectrum from \qty{5}{\GHz} to \qty{8.2}{\GHz}. (b) A Zoom-in spectrum from \qty{6.16}{\GHz} to \qty{6.22}{\GHz} for direct comparison between simulation, modeling and measurement results of Res. 6. (c) Close-up of raw data and fitted measurement data in complex plane. (d) Spectrum with fine frequency span around the resonance frequency of Res 6.}
    \label{fig:spectrum_res5}
\end{figure}

The transmission spectrum, as shown in Fig.~\ref{fig:spectrum_res5}(a), is measured at a power level of around \qty{-100}{dBm} at the sample from \qty{5}{\GHz} to \qty{8.2}{\GHz}. We can identify the 10 dips in transmission to the corresponding resonators. An additional dip occurs at approximately \qty{7.6}{GHz}, which corresponds to an unwanted resonance attributed to impedance mismatch at the interface where the planar CPW line transitions to the top-grounded CPW line.

We extract the resonance frequency, loaded and coupling Q-factors using fine sweeps around the resonance by fitting the transmission to the complete model \cite{Probst_2015, Sage_2011}:
\begin{equation}\label{S21_notch_noise}
S_{21}^{notch}(f) = ae^{j\alpha}e^{-j2\pi f\tau}\left[1-\frac{(Q_l/Q_c)e^{j\phi}}{1+2jQ_l(f/f_r-1)}\right],
\end{equation}
which includes environmental effects such as attenuation a, electric delay $\alpha$ and $\tau$ introduced by the environment and RF cables. Additionally, the wire bonding on the feedline between the chip and the carrier PCB introduces impedance mismatches \cite{Khalil_2012}, which can be observed in Fig.~\ref{fig:spectrum_res5}(c) and are included in (\ref{S21_notch_noise}) by a factor $\phi$.

Using the curve-fitting algorithm from \cite{Probst_2015}, the resonance frequency ($f_r=$ \qty{6.165}{GHz}) and the coupling Q-factor ($Q_c \approx 19.9 \times 10^3$) are extracted, as shown by the fitted curve in black in Fig.~\ref{fig:spectrum_res5}(b). Note that the VNA output power (\qty{-20}{dBm}) is carefully adjusted to ensure sufficient signal-to-noise ratio (SNR) during the measurement, as required for reliable extraction of $f_r$ and $Q_c$ \cite{Probst_2015}. Subsequently, the transmission response of Res. 6 under perfect conditions with the obtained parameters is restored, depicted as the green curve in Fig.~\ref{fig:spectrum_res5}(b). For direct comparison, the simulated $S_{21}$ spectrum and the calculated spectrum from the model for Res. 6 are also included. The results show a frequency downshift of approximately $\Delta f_r \approx$ \qty{60}{MHz} and a coupling $Q_c$ discrepancy of around $\Delta Q_c \approx 10 \times 10^3$.

\begin{figure}[t]
    \centering
    \begin{overpic}{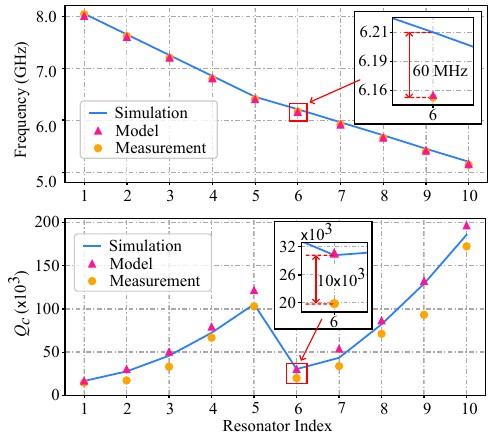} 
        \put(-3, 86){\normalsize (a)}  
        \put(-3, 42){\normalsize (b)}
    \end{overpic}
    \caption{The resonance frequencies (a) and the coupling factors (b) determined from the measurement, simulation and the proposed Model.}
    \label{fig:mea_vs_sim_vs_cal}
\end{figure}

After applying the curve-fitting algorithm to all ten resonators, the determined resonance frequencies and the coupling Q-factors from the measurement are plotted in the Fig.~\ref{fig:mea_vs_sim_vs_cal}. Compared to the simulations, the measurements show a downshift in the spectrum and align more closely with the proposed model. This behavior is typical for superconducting resonators, as the kinetic inductance is not accounted for the simulations. The kinetic inductance is influenced by several factors, including the magnetic penetration depth of the thin-film superconductor ($\lambda_m$), which is determined by the material properties, and the current distribution across the transmission line, which depends on the type and geometry of the transmission line. For a total current flow $I$ in the system, the kinetic inductance can be expressed as \cite{Keiji1992}:
\begin{equation}\label{kinetic_inductance}
L_{l}^{k} = \frac{\mu_0 \lambda_m^2}{|I|^2} \cdot \int J_z^2 dS,
\end{equation}
where $\mu_0$ is the vacuum permeability and $\int dS$ denotes the surface integral of the current density over the conductor cross section.

In this experiment, a thin-film Nb with a thickness of \qty{150}{nm} is deposited on both the bottom and the top substrates. Based on the model proposed by \cite{Gubin2005} and \cite{Pinto_2018}, the magnetic penetration depth is estimated to be approximately $\lambda_m \approx$ \qty{80}{nm}. The current distribution of the top-grounded CPW transmission line can be resolved using cross-sectional simulations in the Ansys Q2D Extractor. By incorporating this information into (\ref{kinetic_inductance}), the kinetic inductance per unit length for the given geometry in Table~\ref{table:cpw_geometry} is calculated to be $L_l^k=$\qty{12}{nH/m} and it is additive to the total inductance for CPW line. While this effect is not included in the primary modeling approach used for extracting resonance frequencies, the following expression is used here solely to illustrate how kinetic inductance alters the fundamental resonance of a superconducting $\lambda/4$ resonator:
\begin{equation}\label{eigen_frequency_with_KI}
f_{r}' = \frac{1}{4 l_{t} \sqrt{(L_l^k + L_l^g) C_l^g}},
\end{equation}
where $L_l^g$ and $C_l^g$ are the geometrical inductance and capacitance per unit length of the top-grounded CPW line, determined using (\ref{Z0_cpw}). For the transmission line with the given values in Table~\ref{table:cpw_geometry} the corresponding geometrical inductance is $L_l^g=$\qty{388}{nH/m}, resulting in a ratio $L_l^k/L_l^g \approx$ \qty{3.1}{\%}, which agrees well to experimental characterizations. Therefore, an approximately 2\% downshift for resonance frequencies determined by either the proposed model or the simulation should be added on.
\begin{figure}[b!]
    \centering
    \begin{overpic}{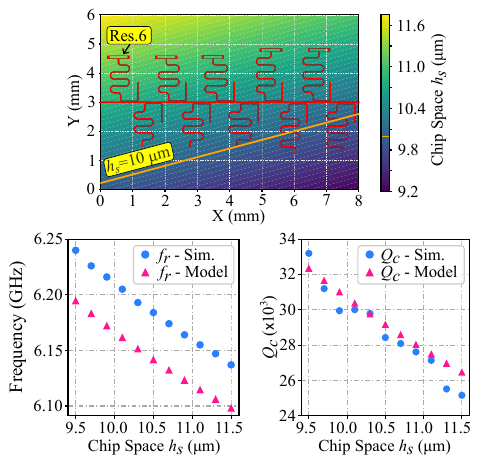} 
        \put(10, 90){\normalsize (a)}  
        \put(0, 44){\normalsize (b)}
        \put(53, 44){\normalsize (c)}
    \end{overpic}
    \caption{(a) Contour map of chip spacing, $h_s$, across the chip relative to the bottom chip. The fitted plane is based on the date directly obtained from optical characterization. The origin (0, 0) is the bottom-left corner where top chip starts. The change of $f_r$ (b) and $Q_c$ (c) when $h_s$ is varying from \qty{9.5}{\micro m} to \qty{11.5}{\micro m}.}
    \label{fig:chip_orientation}
\end{figure}

The $Q_c$ values of the ten resonators extracted using the proposed model show reasonable discrepancies compared to the simulation results as discussed in Section~\ref{subsec:determin_fr_Qc}. However, as shown in Fig.~\ref{fig:mea_vs_sim_vs_cal}(b), larger differences ($\Delta Q_c > 10 \times 10^3$) are observed between the simulation and experimental results. These noticeable discrepancies can be attributed to the control over the chip spacing parameter ($h_s$) during the fabrication process. Although polymer spacers are added to improve the uniformity of $h_s$, the spacing is not consistent across the entire chip. To investigate this, SEM images are taken at specific locations along the chip edges for optical characterization, allowing direct observation of the $h_s$ values. The collected data are fitted, and a contour plot is generated, as shown in Fig.~\ref{fig:chip_orientation}(a). Within the effective region, where the resonators are located, the $h_s$ values are found to vary within the range of \qty{10.5}{\micro m} ± \qty{1}{\micro m}, which is consistent with the tolerance range reported in \cite{Norris_2024}. To quantify the model’s sensitivity to fabrication tolerances, we varied $h_s$ within the standard range of \qty{0.5}{\micro m} ± \qty{1}{\micro m} and evaluated its impact on the determination of $f_r$ and coupling quality factor $Q_c$. Taking Res. 6 as an example, simulations and parameter extractions are performed using the proposed model. As the results are shown in Fig.~\ref{fig:chip_orientation}(b) and (c), the resonance frequency $f_r$ shows a maximum variation of approximately \qty{150}{MHz}, while $Q_c$ is more sensitive to $h_s$, with a maximum variation of about $8\times10^3$. Furthermore, both $f_r$ and $Q_c$ demonstrated a downward trend as $h_s$ increased. As shown in Fig.~\ref{fig:chip_orientation}(a), Res.6 is located at the corner of the chip where $h_s$ is largest. Thus, it is expected that the measured $Q_c$ for Res.6 would decrease significantly compared to both the simulation and the model. Nevertheless, the overall trend of $Q_c$ among the different resonators aligns well with both the simulations and the model results. This suggests that with better control of $h_s$, the experimental results could align more closely with the model predictions.

\section{Conclusion}
This work introduces an analytical model that addresses key limitations in existing approaches to designing feedline-coupled $\lambda/4$ resonators for superconducting quantum circuits. Closed-form expressions for the transmission coefficient $S_{21}$ in two-port measurements are derived, enabling accurate determination of resonator frequency and coupling Q-factors. The model can also be extended to support various chip architectures with minor modifications to the expressions. 

The proposed model is validated through the fabrication and measurement of a test chip featuring resonators with varying geometries, demonstrating good agreement with both FEM simulations and experimental results. The discrepancies in the determined resonance frequencies and coupling Q-factors compared to simulations are within $\Delta f/f_{sim}<$ \qty{1}{\%} and $\Delta Q_c/Q_{c, sim}<$~\qty{20}{\%}, respectively. Despite additional deviations arising from fabrication non-uniformities, the model successfully captures the overall trends in $f_r$ and $Q_c$, confirming its reliability. 

A key advantage of the proposed model is its ability to determine parameters with high speed and minimal computational resources, whereas full-wave EM simulations can take hours on a workstation with 16 cores and 128 GB of RAM. By reducing reliance on computationally intensive simulations, this model facilitates the scalable and efficient design of quantum readout circuits.


%





\ifCLASSOPTIONcaptionsoff
  \newpage
\fi





\bibliographystyle{IEEEtran}
\bibliography{IEEEabrv,Bibliography}

\vfill


\end{document}